\documentclass{ws-jnopm}
\usepackage{mathrsfs}
\usepackage{graphicx}
\usepackage[center]{subfigure}

\begin{document}

\markboth{Guihua Chen, and Shaoqiang Zhang}{Optical solitons in a trinal-channel inverted nonlinear photonic crystal}

%%%%%%%%%%%%%%%%%%% Publisher's Area please ignore %%%%%%%%%%%%%%%%%%%%%%%
\catchline{}{}{}{}{}
%%%%%%%%%%%%%%%%%%%%%%%%%%%%%%%%%%%%%%%%%%%%%%%%%%%%%%%%%%%%%%%%%%%%%%%%%%

\title{Optical solitons in a trinal-channel inverted nonlinear photonic crystal}

\author{GUIHUA CHEN$^1$,SHAOQIANG ZHANG$^{1,2}$, MUYING WU$^1$}
\address{$^1$Department of electronic engineering, Dongguan University of Technology, Dongguan, 523808, China\\
$^2$College of Materials Science and Engineering, South China University of Technology, Guangzhou 510640, China\\
Coresponding author: Guihua Chen, email:cghphys@gmail.com}

\maketitle

\begin{history}
\received{(Day Month Year)}
%\revised{(Day Month Year)}
%\accepted{(Day Month Year)}
%\comby{(xxxxxxxxxx)}
\end{history}

\begin{abstract}
Inverted nonlinear photonic crystals are the crystals featuring competition between linear and nonlinear lattices, with minima of the linear potential coinciding with maxima of the nonlinear pseudopotential, and vice versa. In this paper, a new type of inverted nonlinear photonic crystal is constructed by juxtaposing three kinds of channels into a period. These three channels are a purely linear channel, a saturable self-focusing nonlinear channel, and a saturable self-defocusing nonlinear channel. This optical device is assumed to be fabricated by means of SU-8 polymer material periodically doped with two types of active dyes. The nonlinear propagation of a light field inside this device (passing along the channel) can be described by a nonlinear Schr\"{o}dinger equation. Stable multi-peak fundamental and dipole solitons are found in the first gap of the system. These solitons sufficiently exhibit some interesting digital properties, which may have potential in optical communications.
\end{abstract}


\section{Introduction}

The dynamics of discrete systems is one of research hotspots in optics. Arrays or lattices of evanescently coupled waveguides are prime examples of discrete systems in which discrete optical dynamics can be observed and investigated \cite{Lederer,Christodoulides,ZChen}. Optical fields propagating in such discrete systems exhibit novel phenomena \cite{YVK1,schwartz,Joushaghani,Trompeter}. However, traditional coupled arrays or lattices, such as AlGaAs waveguide arrays \cite{Eisenberg}, periodically poled lithium niobate waveguides \cite{Iwanow}, photorefractive crystals \cite{Fleischer,Efremidis,Yongyao1}, and liquid crystals \cite{Fratalocchi} via optical induction, are almost composed of by passive materials. It is well known that active materials (AMs) have plenty of optical characteristics, such as the strong dispersion, the complex dielectric constant, and the strong variation of the dispersion relation near the resonance. Therefore, waveguide arrays made by the active materials show advantages of freedom to manipulate, and offer low threshold, in comparison with their passive counterparts. These properties qualify the active waveguides to have many potential optical applications, such as soliton or bullet generation \cite{Yingwu,GHuang,Ghuang2,Thong,CHuang,Yongyao4,Weipang,JWu,JGao,Blaauboer}, optical switching \cite{Prineas}, optical storage \cite{Toader,JYZhou,Melnikov,Khomeriki}, PT-symmetry formation \cite{Hu1,Hu2,Makris1,Makris2,Nixon,Guo}, and nonlinear optical frequency conversion \cite{Juntao1,Yongyao2}.

Recently, a 2D imaginary part photonic crystal (IPPhC) (or resonantly absorbing waveguide arrays, RAWA) has been realized by the technique of multi-beam interference holography lithography and back-filling \cite{Juntao,Mfeng,Bliang}, which allows the active materials to be doped into a homogeneous background with spatial structure. The active material of this IPPhC is chosen to be the Rhodamine B (RhB, a dye featuring saturable absorption), which can be doped in a homogeneous SU-8 background to form the IPPhC \cite{Bliang}. In the low power region, where the nonlinearity of the active material is not excited, the IPPhC shows a significant efficiency on color-seperation \cite{Yikunliu} , which may have a potential to construct a new kind of color filter matrix in display technology and industry. In the high power region, where the nonlinearity of the active material cannot be neglected, the nonlinear localized modes, i.e. solitons, show some interesting power-dependent properties. The centers of the solitons can switch between the linear and nonlinear stripe as the power is increasing \cite{yongyao3}.

In this paper, we introduce a kind of one-dimensional discrete optical system, which is developed from the color filter matrix in Ref. \cite{Yikunliu}  and consider it within the high power region. This system is assumed to be built on the basis of the SU-8 doped with two kinds of active materials (alias, two kinds of dyes) featuring saturable absorption peak in different wavelength respectively. In Section II, a description to this model is presented in detail, and a nonlinear Schr\"{o}dinger equation, which describe the nonlinear propagation of the light field though the system, is drawn in the paraxial approximation. In section III, a numerical simulation is carried out. Multi-peak fundamental and dipole solitons, which show many interesting digital properties, are found in the first band gap of the system. And the paper is concluded in Section IV.

\begin{figure}[htbp]
\centering%
{\includegraphics[scale=0.8]{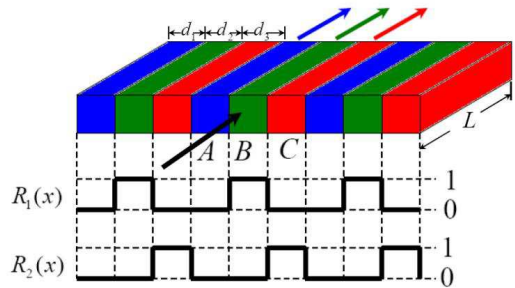}}
\caption{The diagram of 1-D optical device. The green and red areas are doped with AMs, and the blue ones are the background media. $R_1(x)$ and $R_2(x)$ are the structure functions of the two AMs. $d_1\rightarrow d_3$ are the widths of the channels A-C, respectively.}
\label{Fig1}
\end{figure}

\section{The model}

The structure of our device is displayed in Fig. \ref{Fig1}. The propagating light field through the device can be described (in the paraxial approximation) with the underlying equation:

\begin{eqnarray}
i{\partial E\over\partial z}=-{1\over2k}{\partial^{2} E\over\partial x^{2}}-\sum^2_{j=1}{(k_0 \Delta n_j +{i\alpha_j\over2})R_j(x)\over1+|E|^2}E \label{NLS}
\end{eqnarray}

In Eq. (\ref{NLS}), $k=k_0n$, where $n$ is the refractive index of the background, $k_0$ is the vacuum vector. $\alpha_j$ ($j=1,2$) are the absorption coefficients of the two AMs, respectively. $\Delta n_j$ ($j=1,2$) are the corresponding refractive index difference (RID) induced by the two absorption coefficients, respectively. The relation between the absorption coefficient and the correspondent RID is assumed to be satisfied by the Kramers-Kronig (K-K) relationship. $R_j(x)$ ($j=1,2$) are the two structure functions of the AMs. $R_j(x)=1$ describes the areas doped with AM $j$. In Fig. \ref{Fig1}, AM1 is doped into the green stripes while AM2 is doped into the red ones,  and blue stripes in the figure represent the background (alias, pure SU-8). These three channels are juxtaposed into one period.

\begin{figure}[htbp]
\centering%
\subfigure[] {\label{Fig2a}
\includegraphics[scale=0.25]{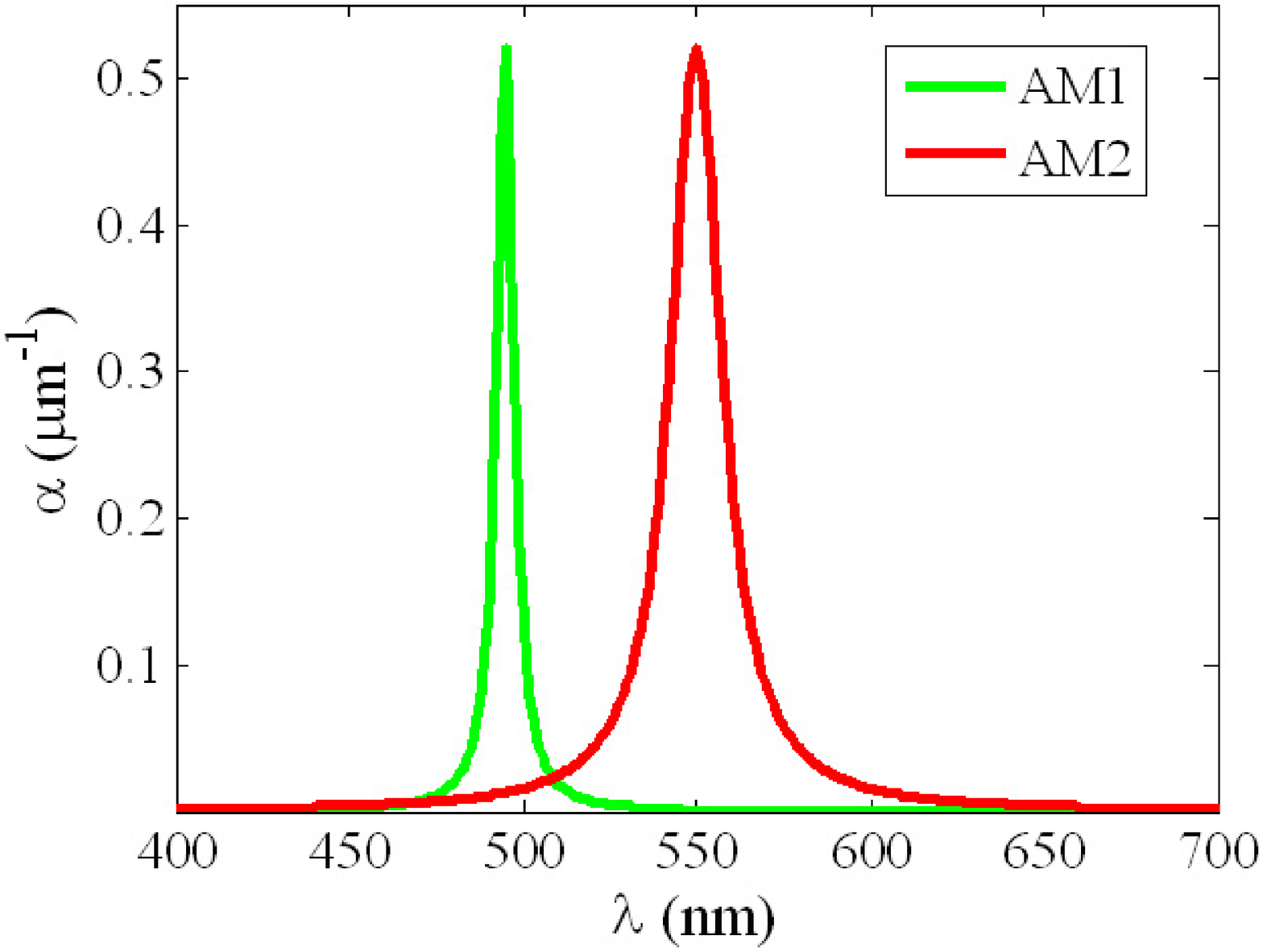}}
\subfigure[] {\label{Fig2b}
\includegraphics[scale=0.295]{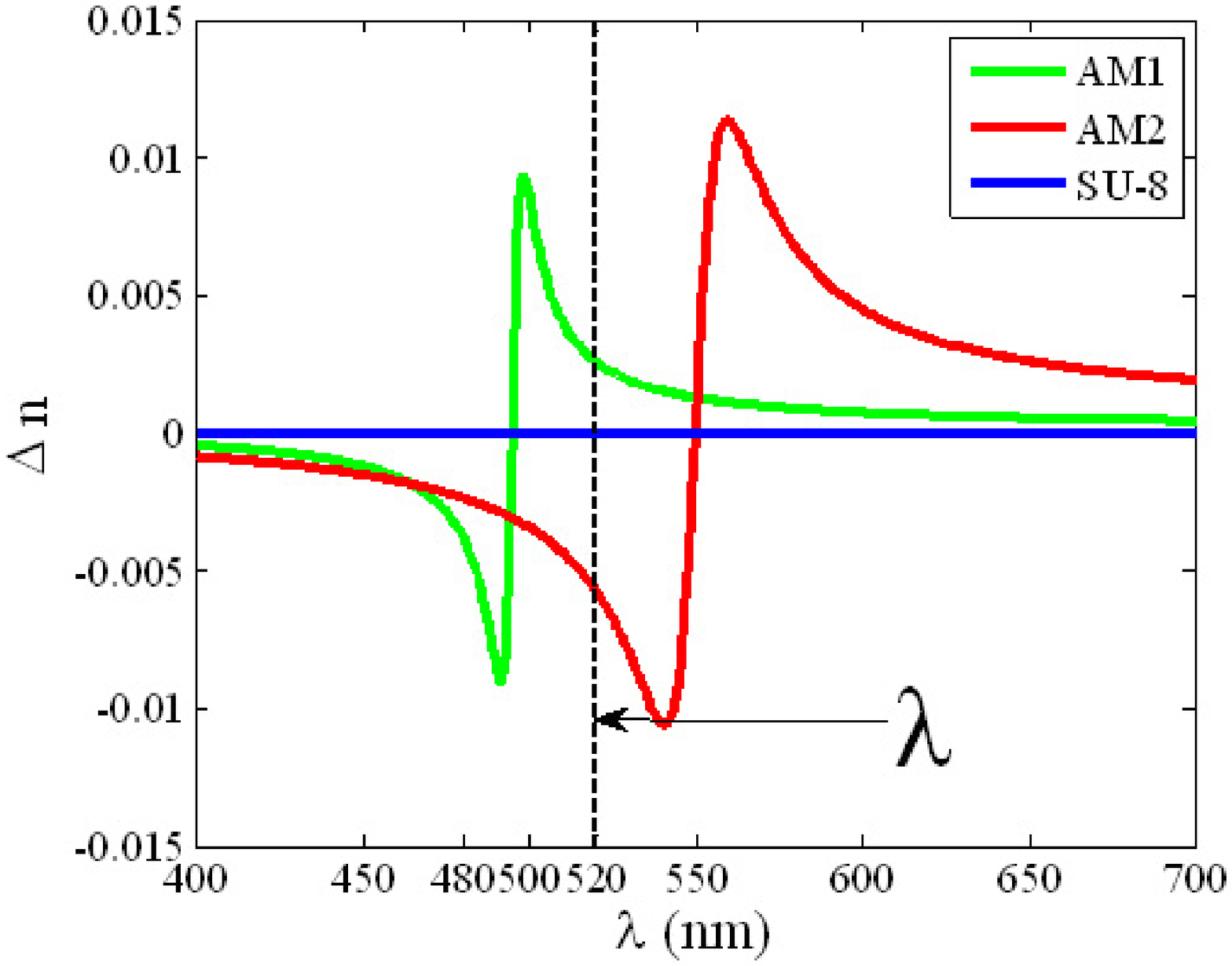}}
\caption{(Color online) (a) The absorption curves of the two AMs. (b) The refractive index difference curves of two AMs. The arrow is the wavelength of the light we chosen for incidence the device. }
\label{Abs_co}
\end{figure}

We assume that the absorption coefficients of the two AMs have a Lorentzian shape, which are shown in FIG. \ref{Fig2a} (We assume the absorption centers of them are $\lambda=480$ nm and $560$ nm, respectively). According to the K-K relation, their RIDs are figured out and shown in FIG. \ref{Fig2b}.

In theory, there may exist such a possibility: at the region between the two absorption peaks of the AMs (See in Fig. \ref{Fig2b}, i.e. $\lambda$=520 nm), we obtain the RID of the two AMs higher and lower the background, respectively. By neglecting the absorption coefficients at this wavelength, Eq. (\ref{NLS2}) can be simplified as:

\begin{eqnarray}
iE_z=-{1\over2k}E_{xx}+\sum^2_{j=1}{K_j(x)\over1+|E|^2}E \label{NLS2}
\end{eqnarray}
where $K_j(x)=-k_0\Delta n_j(\lambda)R_j(x)$ ($j=1,2$).

Figure \ref{Fig2b} shows that, for AM1, $\Delta n_1(\lambda)>0$, while for AM2, $\Delta n_2(\lambda)<0$. For convenience, if we further assume $|\Delta n_2(\lambda)|=|\Delta n_2(\lambda)|=\kappa$, Eq. (\ref{NLS2}) can be simplified as following dimensionless nonlinear Schr\"{o}dinger equation:

\begin{eqnarray}
iU_z=-{1\over2}U_{xx}-{V_1(x)\over1+|U|^2}U+{V_2(x)\over1+|U|^2}U \label{NLS3}
\end{eqnarray}
where $U$ is a dimensionless light field; $V_j(x)=V_0R_j(x)$ ($j=1,2$) with $V_0=k_0\kappa$; $k$ is removed by rescaling of $x$. From Fig. \ref{Fig1}, we know that the light field will experience saturable self-focusing(SF) in channel C (Red color stripe) and saturable self-defocusing(SDF) in channel B (Green color stripe), while in channel A (Blue color stripe), the field is linear propagation.

\begin{figure}[htbp]
\centering%
{\includegraphics[scale=0.3]{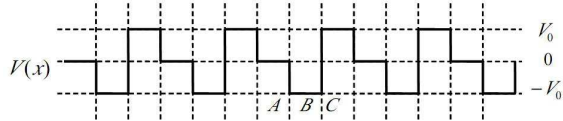}}
\caption{Total linear potential of Eq. (\ref{NLS3})}
\label{Fig3}
\end{figure}

If we take off the saturable nonlinearity from Eq. (\ref{NLS3}), the total linear potential $V(x)=V_1(x)-V_2(x)$ converts into a one-dimensional periodical potential with a fine structure. By considering the nonlinearity, the modulation to the nonlinear potential is exactly equal to $-V(x)$, which forms a combination of competing $\pi$-out-of-phase juxtaposed linear potential and nonlinear potential. The maxima of the refractive index coincides with the maxima of the local strength of the self-defocusing nonlinearity (the bottom of the linear potential coincide with the top of the nonlinear potential), while the minima of the refractive index coincides with the maxima of the self-focusing nonlinearity (the top of the linear potential coincide with the bottom of the nonlinear potential). The medium of this type is naturally called an inverted nonlinear photonic crystals (INPCs) \cite{YVK2,YVK3,Yongyao5}. However, previous INPCs mainly contain one type of nonlinearity (self-focusing or defocusing); in our model, the mixture of two types of nonlinearity is considered.

In the following section, we will carry out the numerical simulations on Eq. (\ref{NLS3}) and find that this system can afford the existence of a rich variety of multi-peak solitons. The numbers of the peaks in the soliton solutions show interesting digital properties, which may have potential in developing new types of optical devices.

\section{Numerical Results}

We assume that the stationary solutions of Eq. (\ref{NLS3}) can be described with $U(x,z)=u(x)e^{-i\mu z}$, where $-\mu$ is the propagation constant of the soliton. The stability of the stationary solitons can be numerically identified by the computation of eigenvalues for small perturbations or direct simulation. For the eigenvalue problem, the perturbed solution is given as $U=(u+we^{i\lambda z}+v^{\ast}e^{-i\lambda^{\ast}z})e^{-i\mu z}$. Substitution of this ansatz into Eq. (\ref{NLS3}) and linearization can lead to the eigenvalue problem
\begin{equation}
\left[
\begin{array}{cc}
{1\over2}{d^{2}\over dx^{2}}+\mu-\sum_{i}{V_{i}(x)\over F}\left(1-{u^{2}\over F}\right) &
\sum_{i}{V_{i}(x)\over F}u^{2} \\
\sum_{i}{V_{i}(x)\over F}u^{2} & -{1\over2}{d^{2}\over d x^{2}}-\mu+\sum_{i}{V_{i}(x)\over F}\left(1-{u^{2}\over F}\right)
\end{array}%
\right] \left(
\begin{array}{c}
w \\
v%
\end{array}%
\right) =\lambda \left(
\begin{array}{c}
w \\
v%
\end{array}%
\right)  \label{lambda}
\end{equation}%
where $F=1+u^{2}$ and $i=1,2$. The solution $u$ is stable if all the eigenvalues of Eq. ({\ref{lambda}}) are real.

We select the Newton-Jocobi method to figure out the soliton solutions \cite{JYang}. In the numerical simulation, we choose the modulation depths of $V_1(x)$ and $V_2(x)$ to be $V_0=0.3$, and the widths of the three stripes to be $d_1=d_2=d_3=2.25$. We find that valid soliton solutions seem mainly converge in the vicinity of zero of the propagation constant $-\mu$. Some multi-peak fundamental and dipole solitons can exist stably.

\begin{figure}[htbp]
\centering%
\subfigure[] {\label{Fig4a}
\includegraphics[scale=0.33,bb= 96 238 515 553, clip=true]{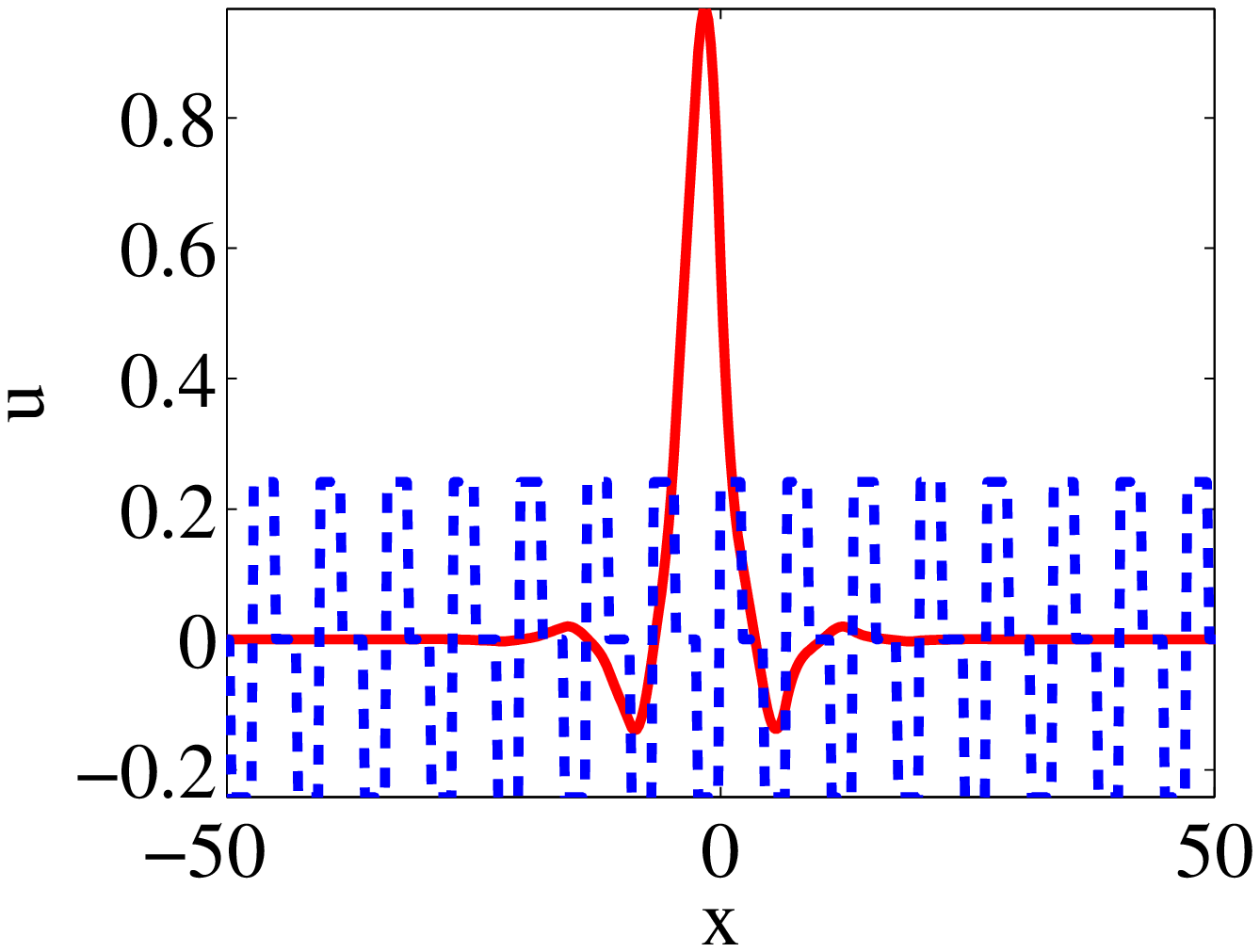}}
\subfigure[] {\label{Fig5a}
\includegraphics[scale=0.33,bb= 96 238 515 553, clip=true]{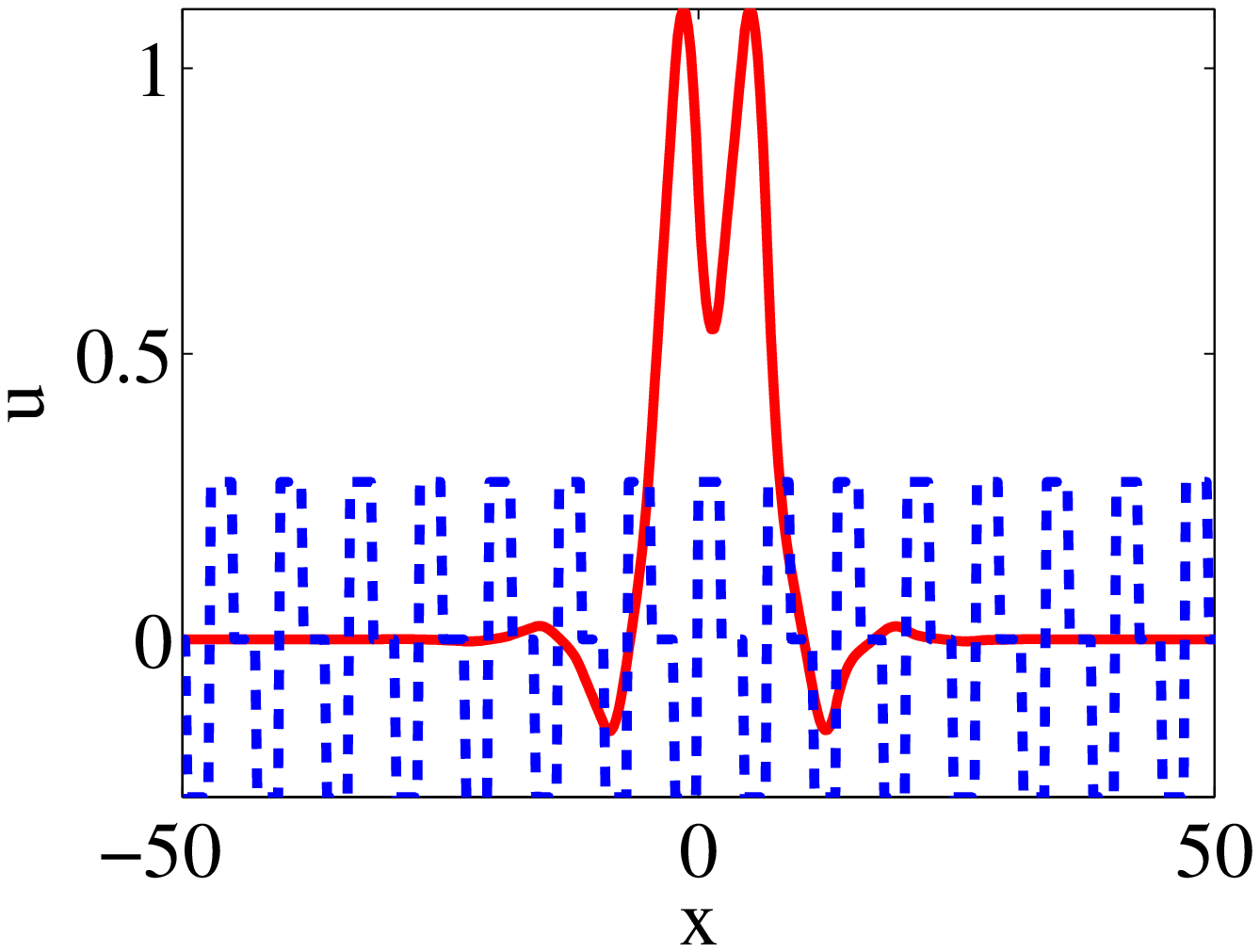}}
\subfigure[] {\label{Fig6a}
\includegraphics[scale=0.33,bb= 96 238 515 553, clip=true]{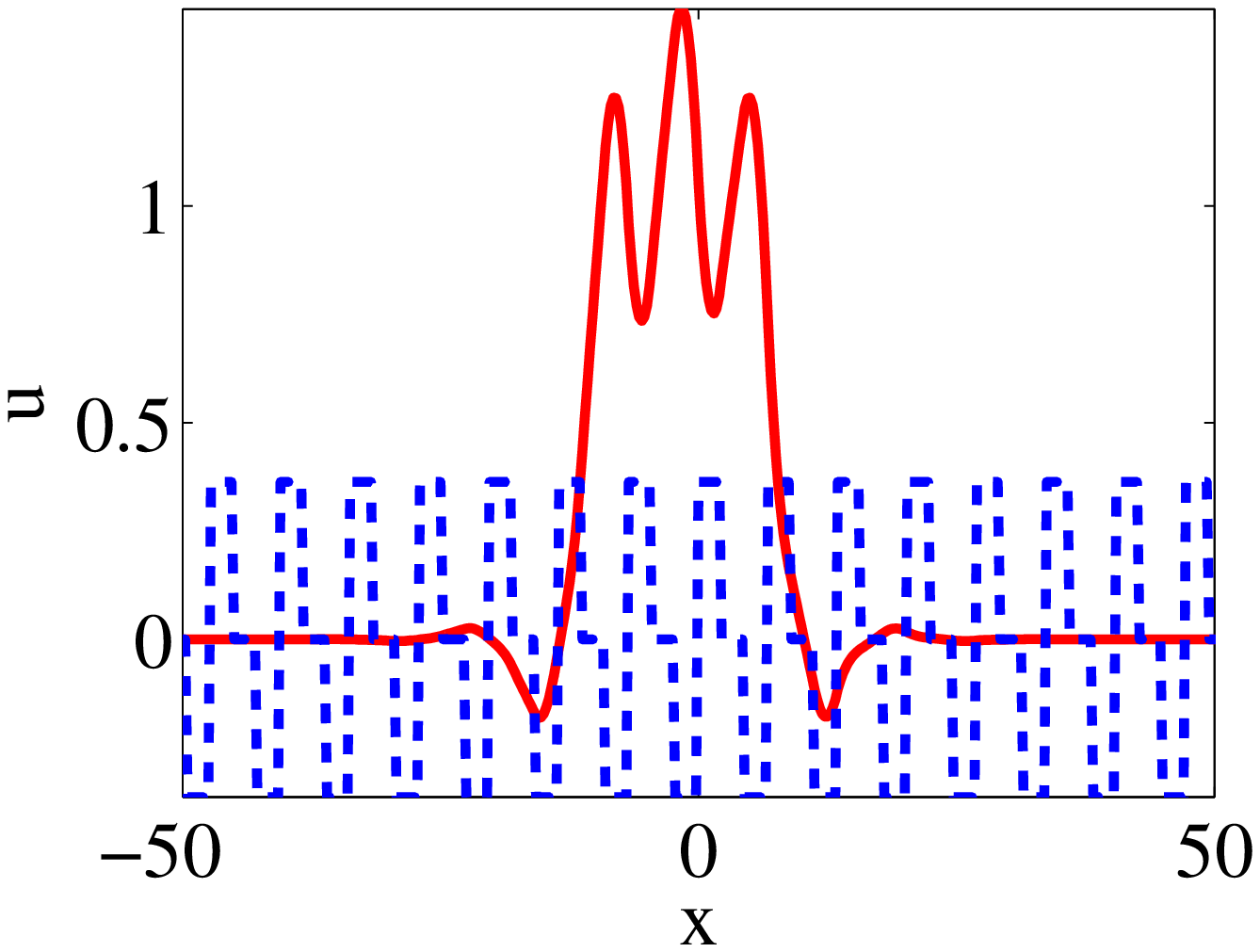}}
\subfigure[] {\label{Fig7a}
\includegraphics[scale=0.33,bb= 96 238 515 553, clip=true]{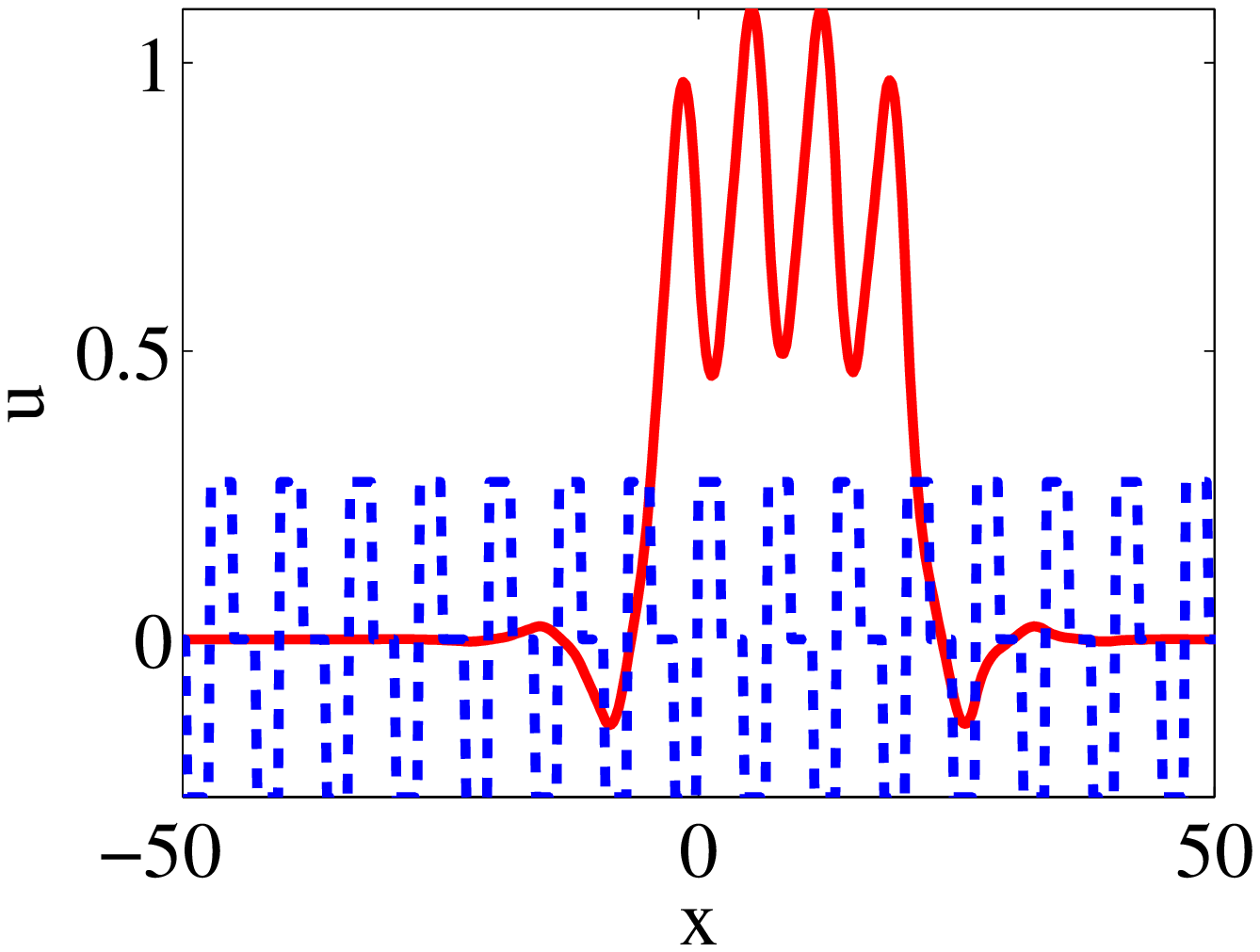}}
\subfigure[] {\label{Fig8a}
\includegraphics[scale=0.33,bb= 96 238 515 553, clip=true]{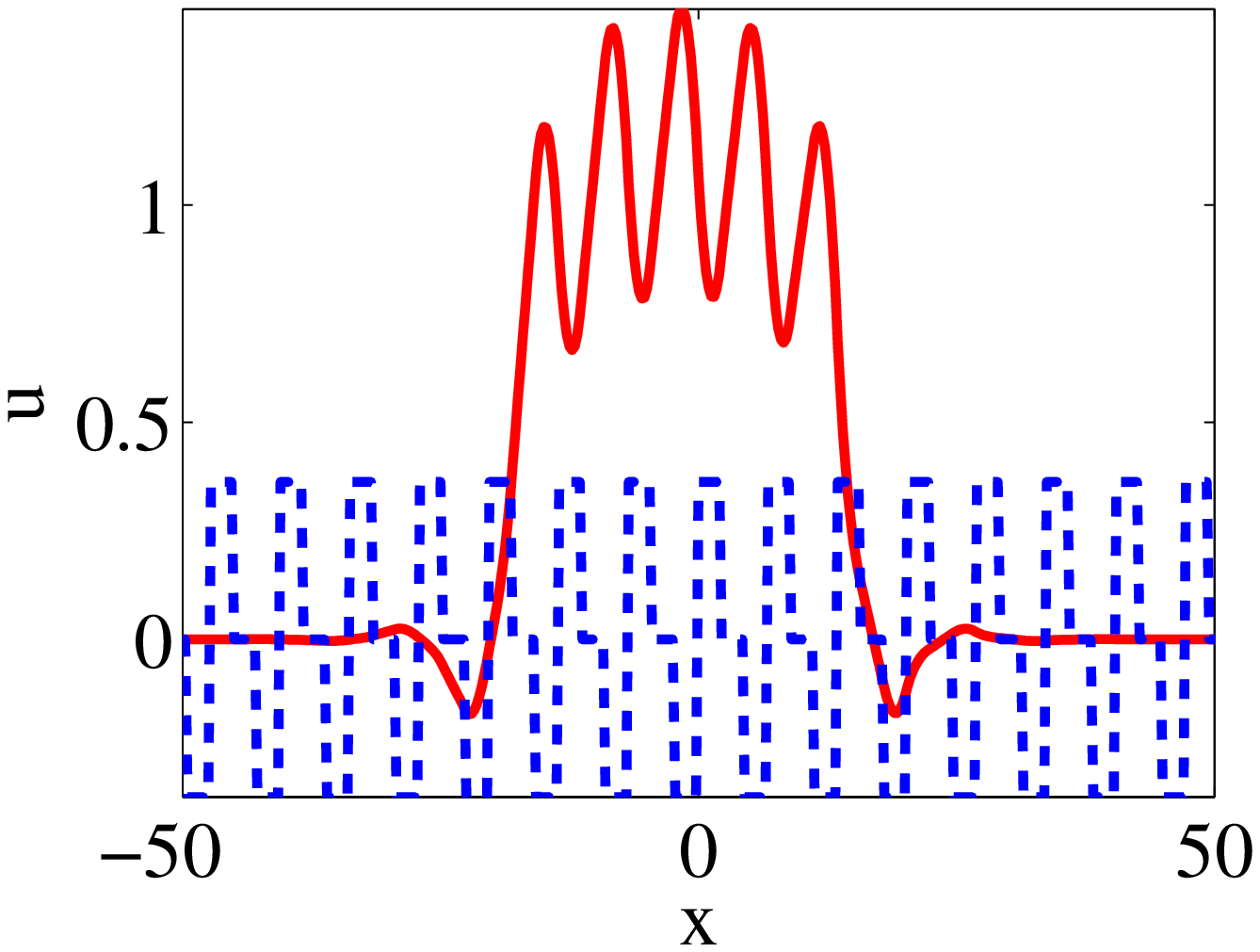}}
\subfigure[] {\label{Fig9a}
\includegraphics[scale=0.33,bb= 96 238 515 553, clip=true]{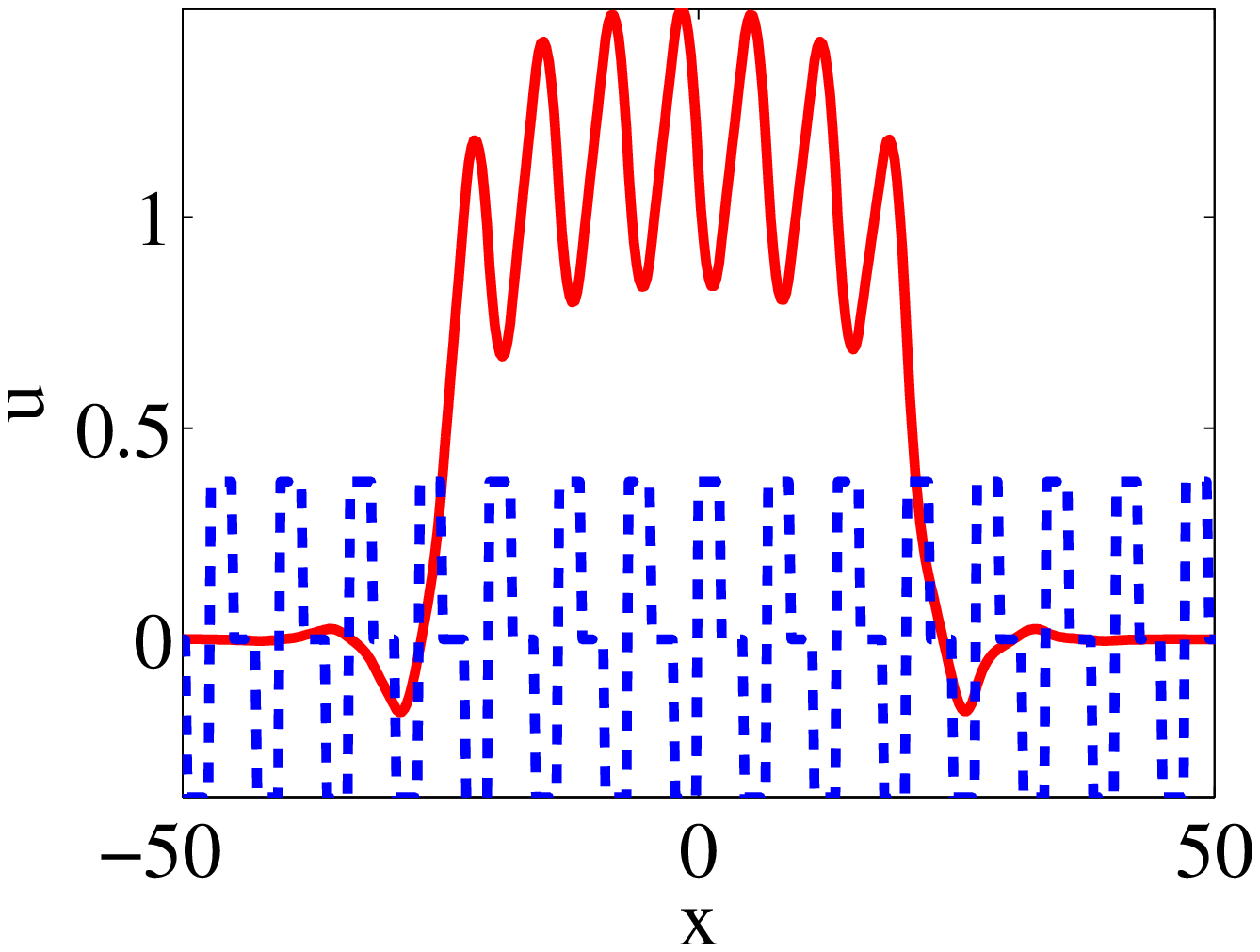}}
\caption{(Color online) The amplitudes of some multi-peak fundamental solitons. (a) 1-peak with $\mu=0.005, P=3.0453$. (b) 2-peaks with $\mu=0.00123, P=8.702$. (c) 3-peaks with $\mu=0.00535, P=20.6291$. (d) 4-peaks with $\mu=-0.0085, P=15.3901$. (e) 5-peaks with $\mu=0.00238, P=35.6062$. (f) 7-peaks with $\mu=0.00239, P=55.1836$.}
\label{fundamental}
\end{figure}

The fundamental solitons with the number of peaks equal to 1, 2, 3, 4, 5 and 7  are shown in Fig. \ref{fundamental}.

\begin{figure}[htbp]
\centering%
\subfigure[] {\label{Fig8c}
\includegraphics[scale=0.33,bb= 0 5 410 553, clip=true]{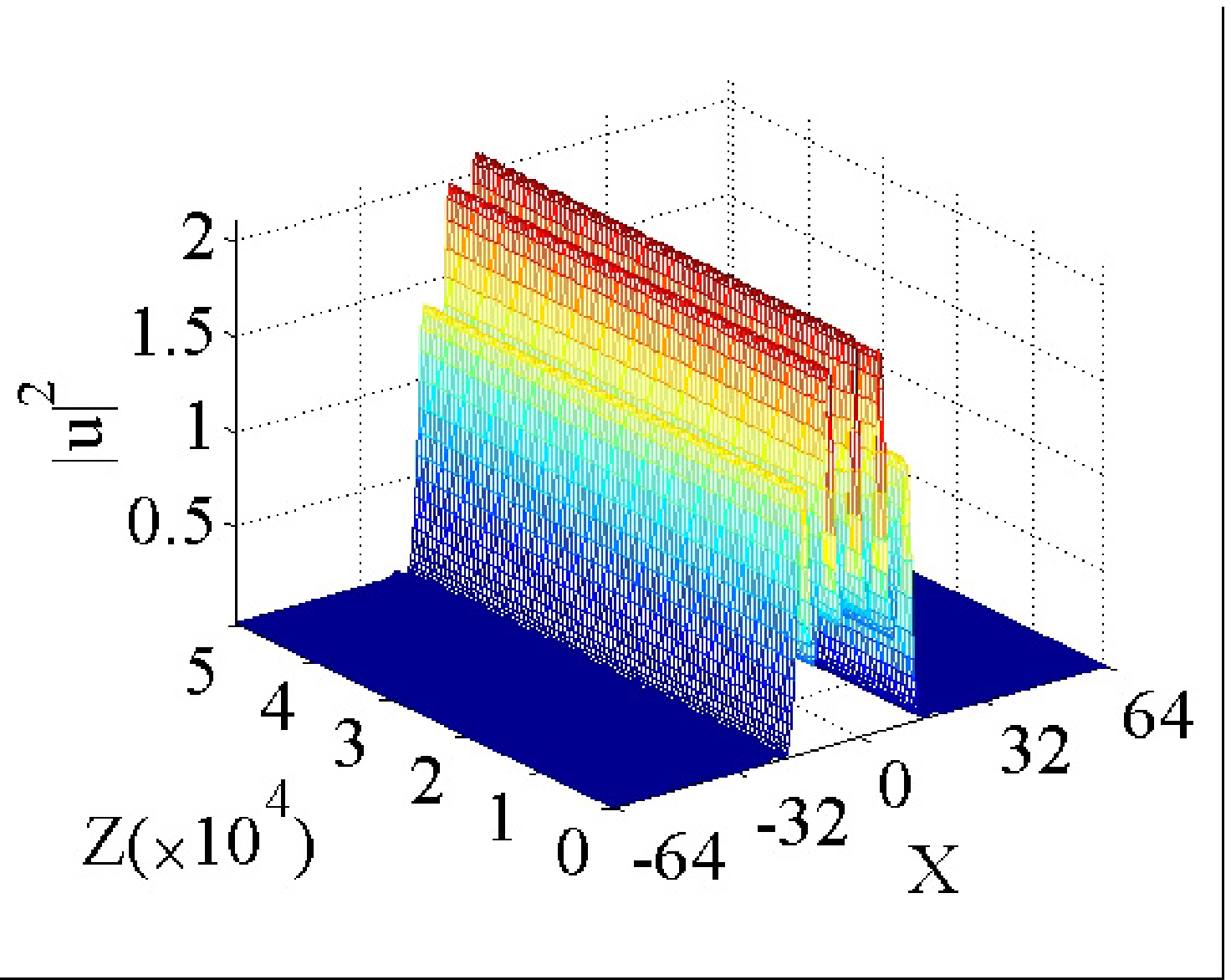}}
\subfigure[] {\label{Fig8d}
\includegraphics[scale=0.33,bb= 96 238 515 553, clip=true]{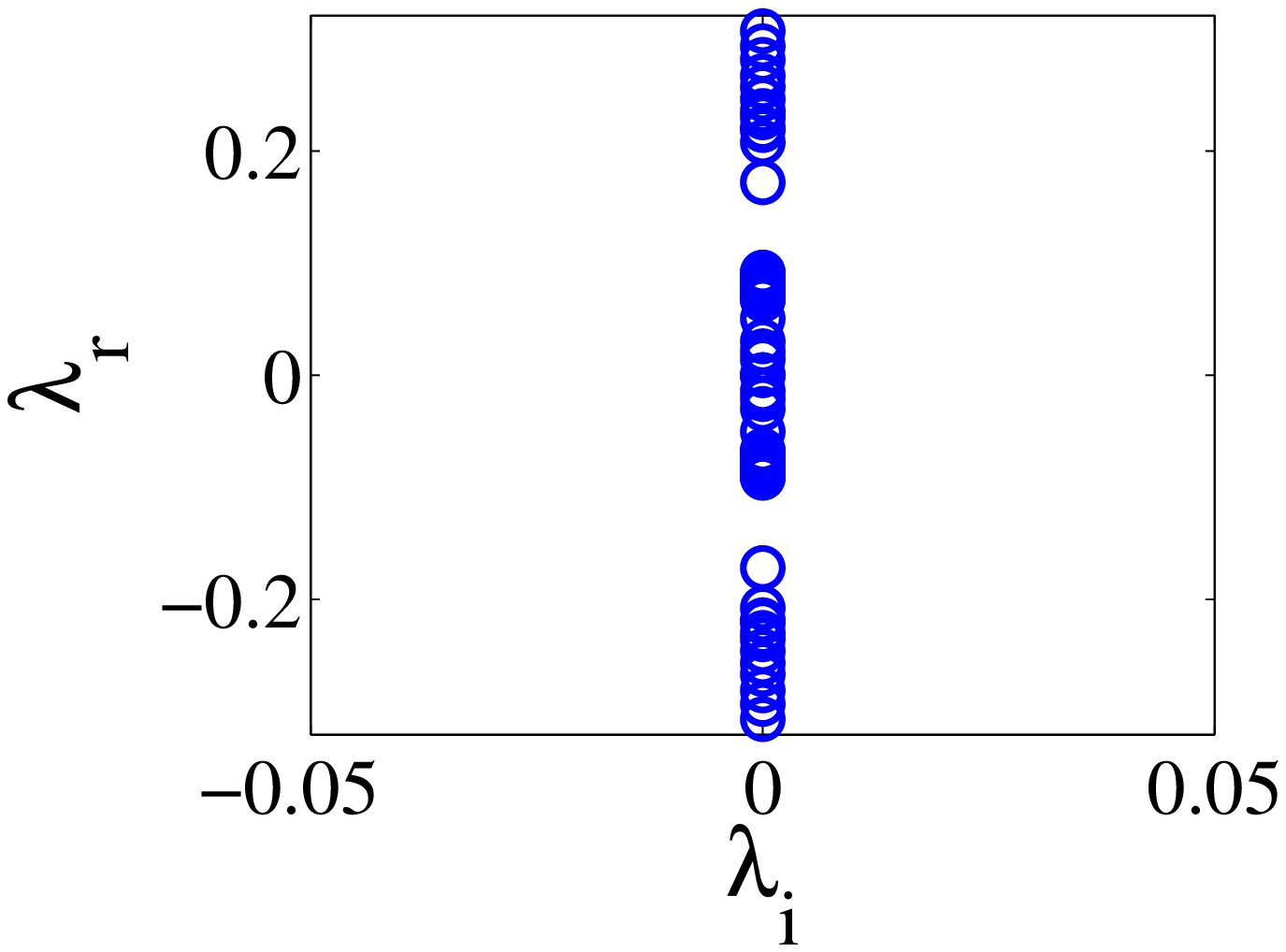}}
\caption{(Color online) The stability of the soliton in Fig. \ref{Fig8a}. (a) The evolution of the soliton perturbed by $1\%$ noises. (b) The growing rate of the soliton.}
\label{stable1}
\end{figure}

The stabilities of these soliton solutions are identified by the computation of eigenvalues for small perturbations and direct simulation. The numerical results show that all of them are stable. Typical example (5 peaks case) of such analysis is shown in Fig. \ref{stable1}.

\begin{figure}[htbp]
\centering%
\subfigure[] {\label{Fig10a}
\includegraphics[scale=0.33,bb= 96 238 515 553, clip=true]{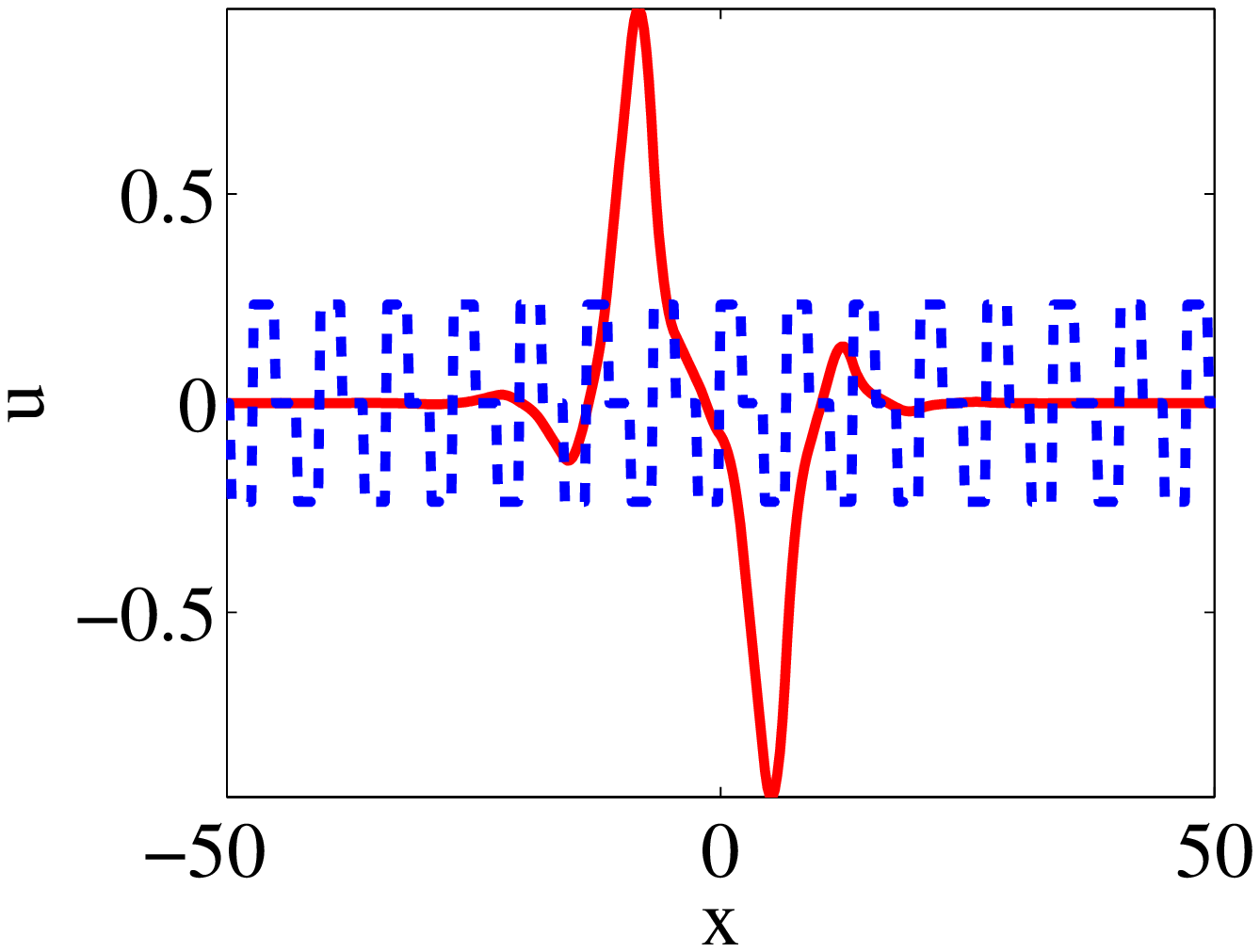}}
\subfigure[] {\label{Fig11a}
\includegraphics[scale=0.33,bb= 96 238 515 553, clip=true]{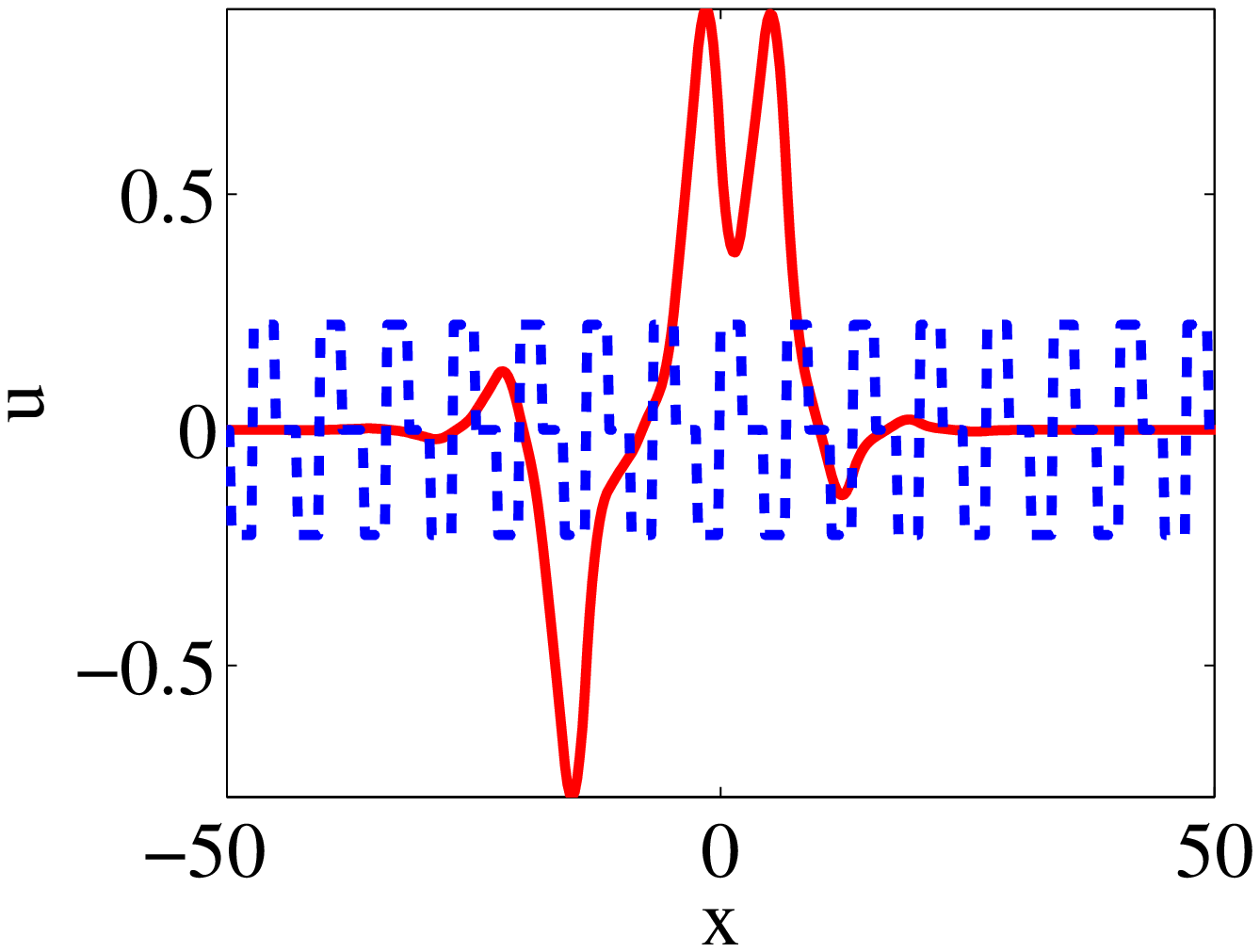}}
\subfigure[] {\label{Fig12a}
\includegraphics[scale=0.33,bb= 96 238 515 553, clip=true]{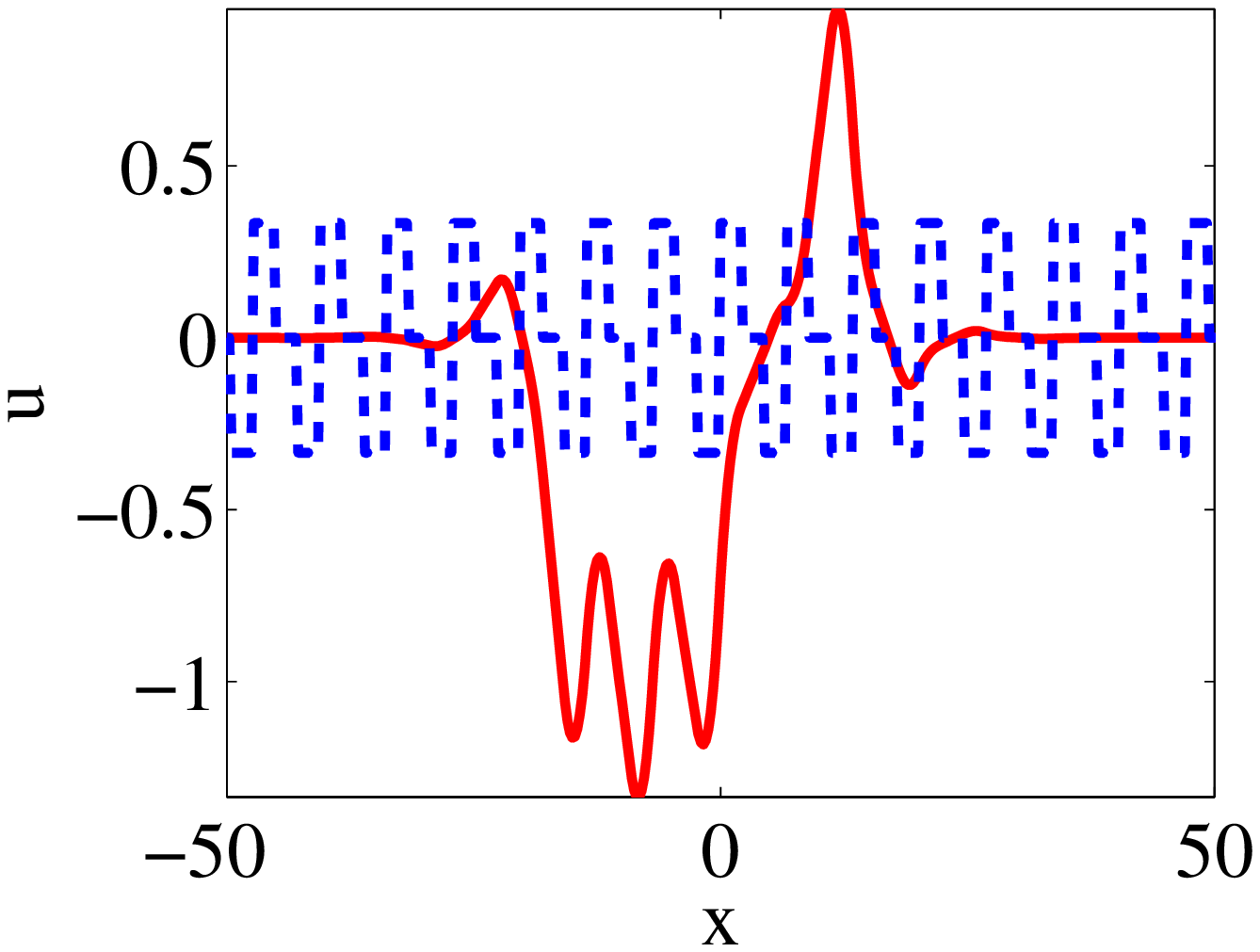}}
\subfigure[] {\label{Fig13a}
\includegraphics[scale=0.33,bb= 96 238 515 553, clip=true]{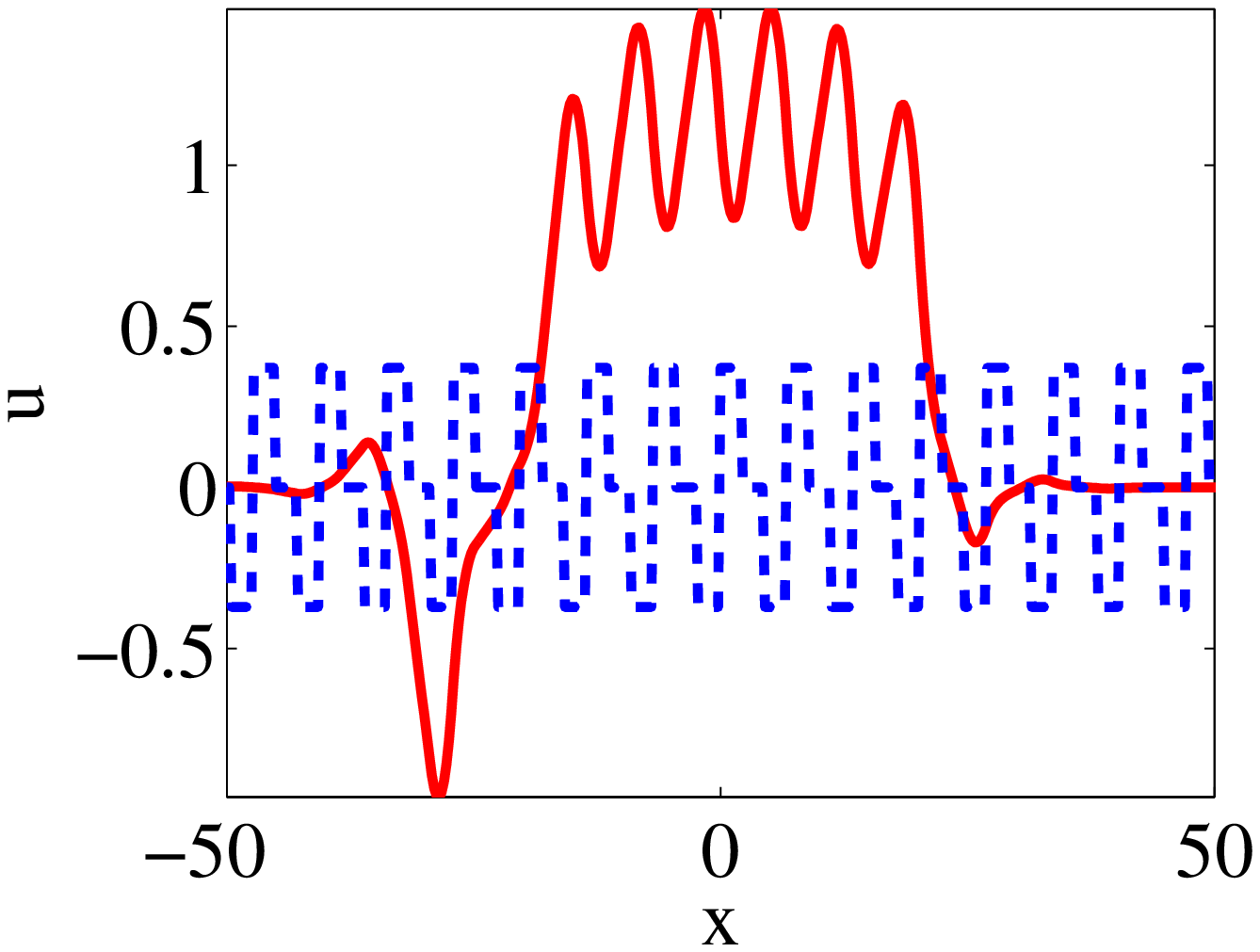}}
\subfigure[] {\label{Fig14a}
\includegraphics[scale=0.33,bb= 96 238 515 553, clip=true]{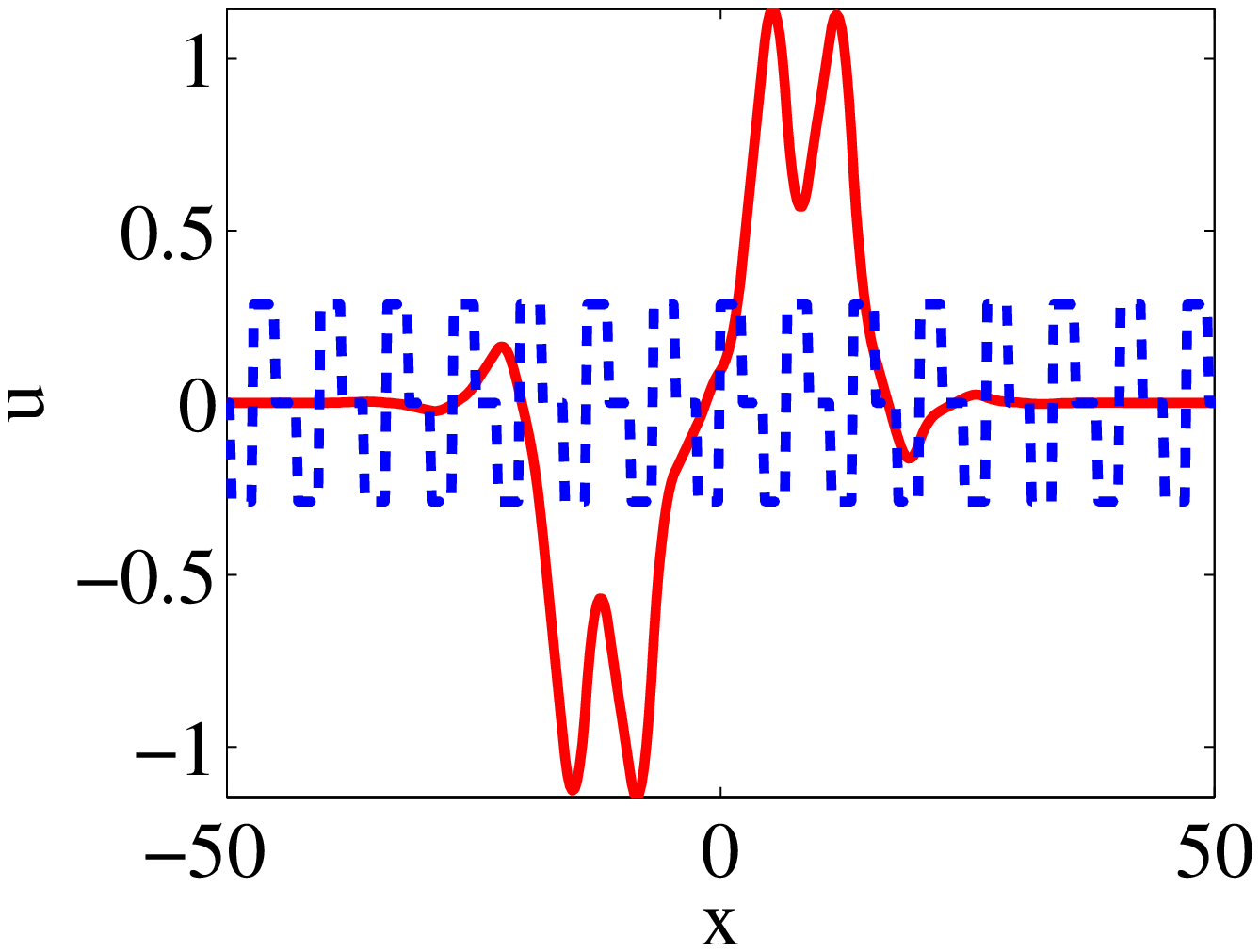}}
\subfigure[] {\label{Fig15a}
\includegraphics[scale=0.33,bb= 96 238 515 553, clip=true]{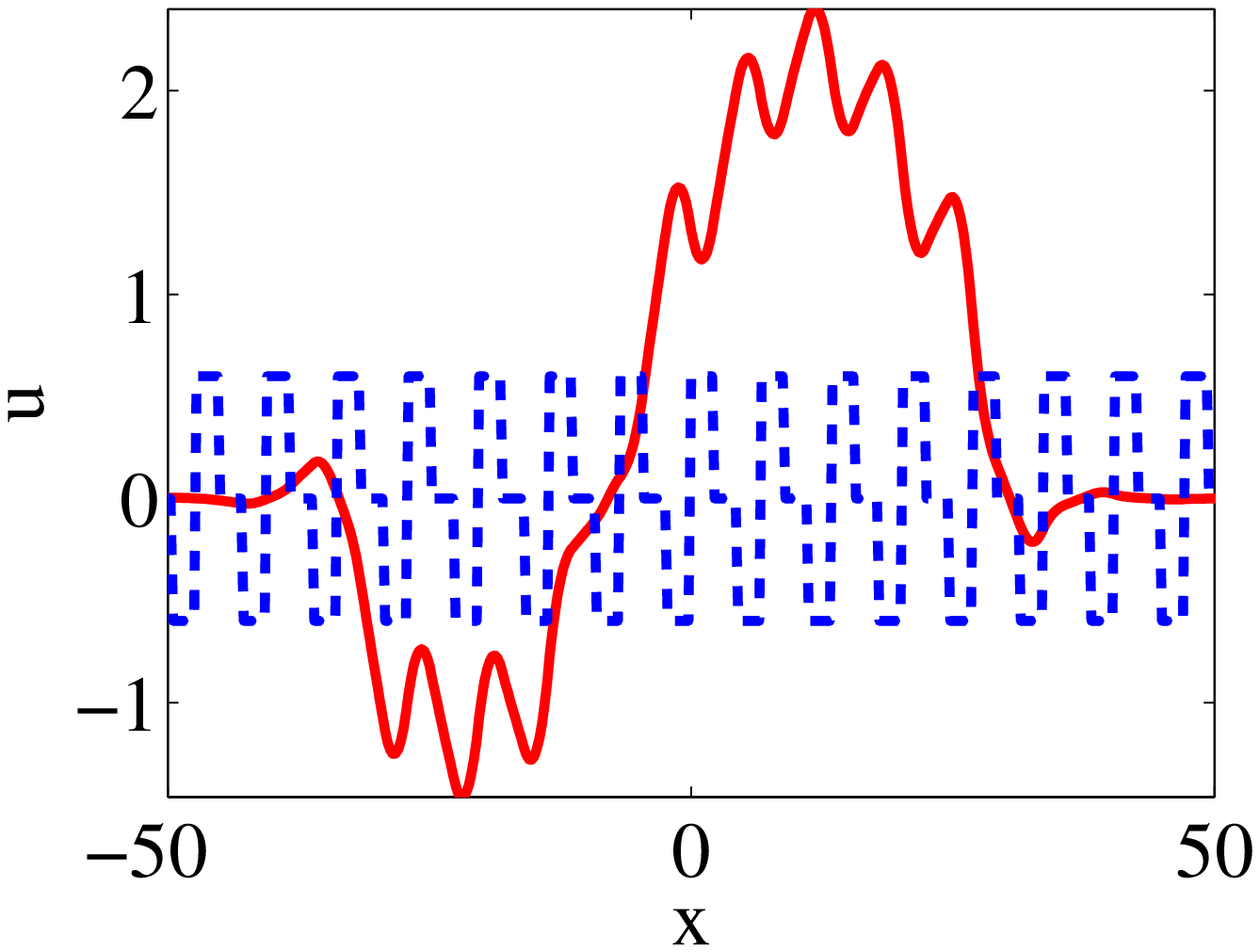}}
\caption{(Color online) The amplitudes of some multi-peak dipole solitons. (a) 1+1 peaks with $\mu=0.00133, P=5.7490$. (b) 1+2 peaks with $\mu=-0.0137, P=7.1972$. (c) 1+3 peaks with $\mu=0.00228, P=20.2986$. (d)1+6 peaks with $\mu=0.00256, P=49.4150$. (e)2+2 peaks with $\mu=0.00232, P=18.6655$. (f)3+5 peaks with $\mu=0.0053388, P=116.8560$.}
\label{dipole}
\end{figure}

\begin{figure}[htbp]
\centering%
\subfigure[] {\label{Fig11c}
\includegraphics[scale=0.33,bb= 0 5 410 553, clip=true]{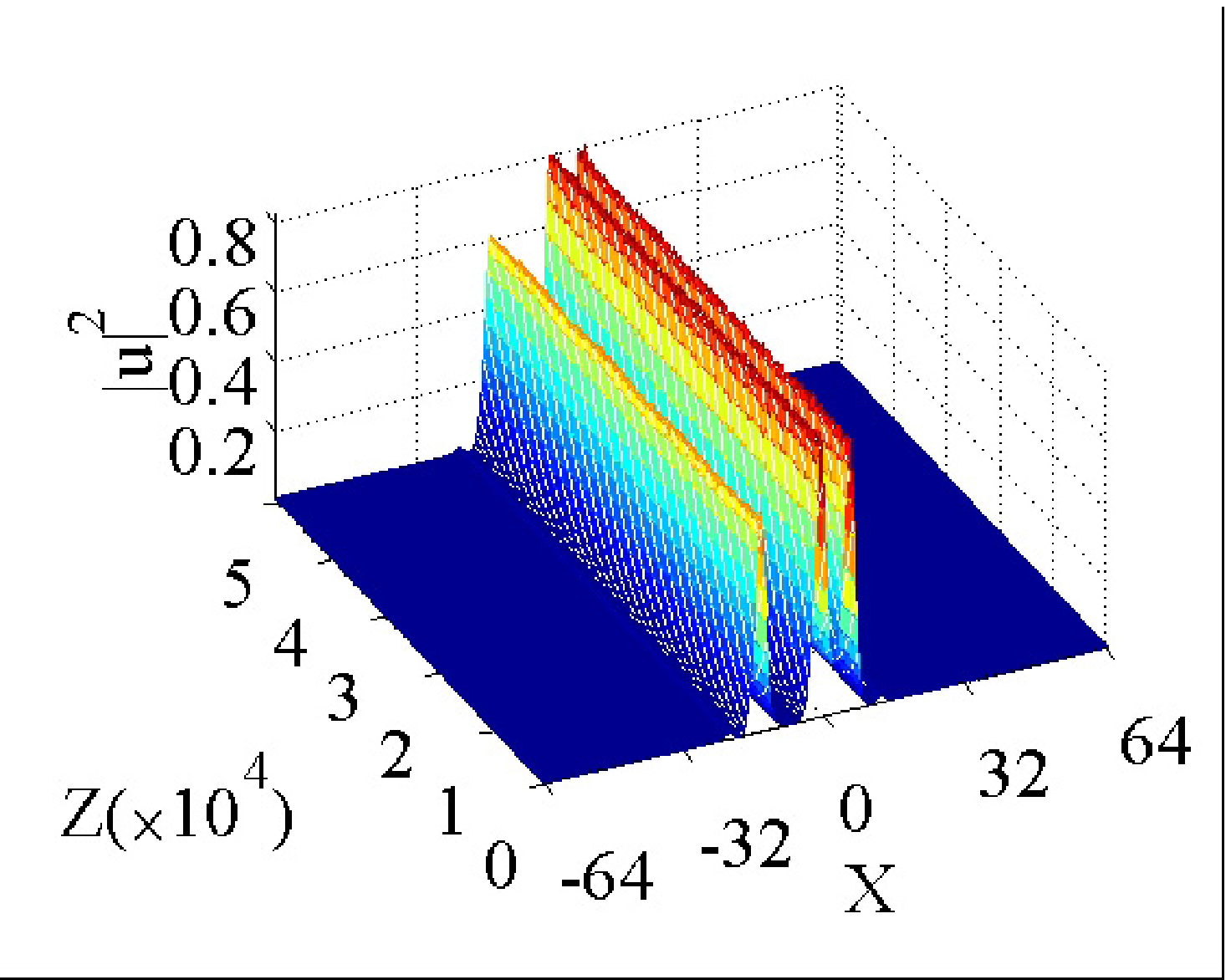}}
\subfigure[] {\label{Fig11d}
\includegraphics[scale=0.33,bb= 96 238 515 553, clip=true]{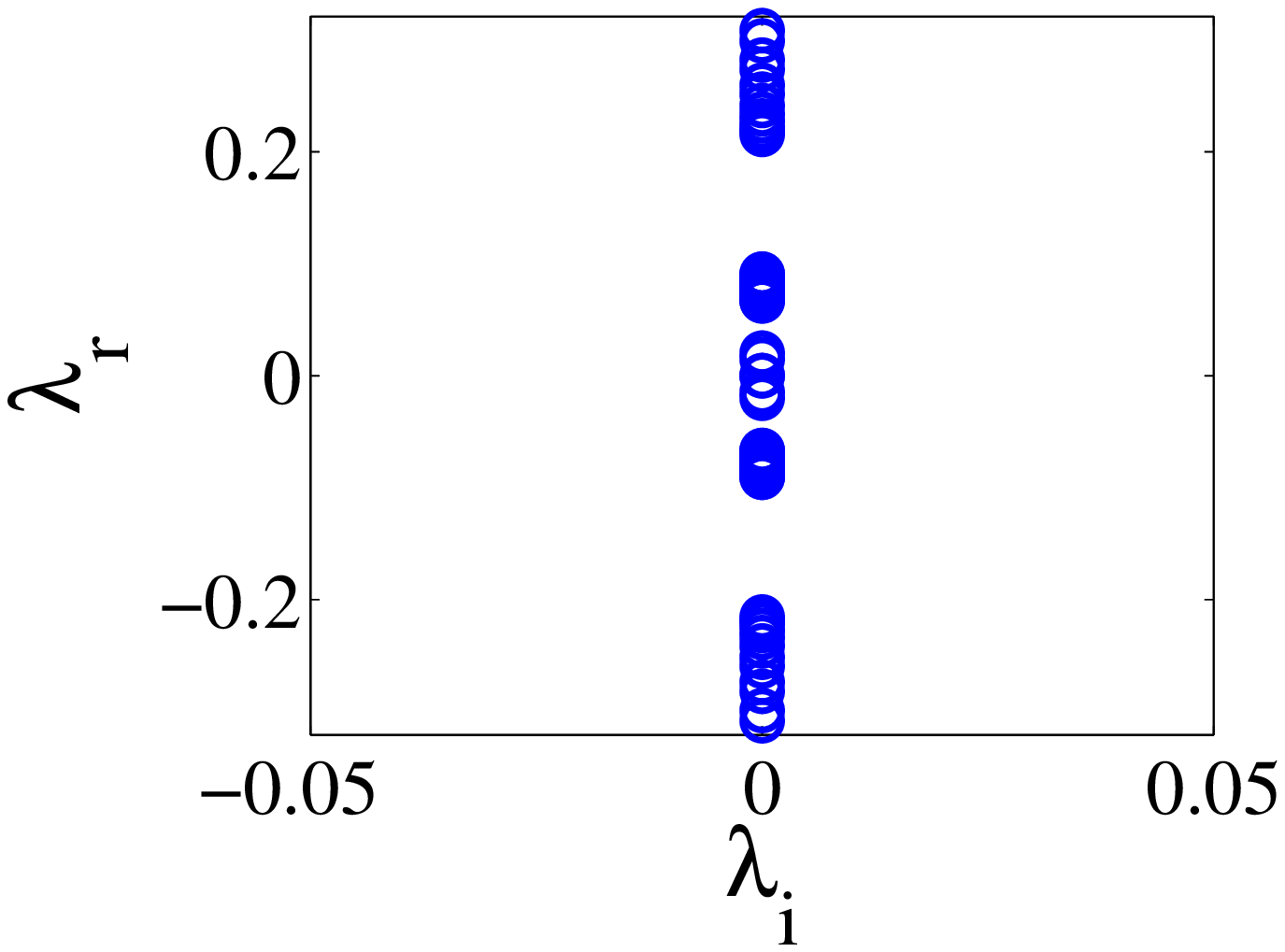}}
\caption{(Color online) The stability of the soliton in Fig. \ref{Fig11a}. (a) The evolution of the soliton perturbed by $1\%$ noises. (b) The growing rate of the soliton.}
\label{stable2}
\end{figure}

Multi-peak dipole solitons can also be found. Figure \ref{dipole} displays some multi-peak dipole solitons with 1+1 peaks, 1+2 peaks, 1+3 peaks, 1+6 peaks, 2+2 peaks and 3+5 peaks. Typical example of the stability analysis is shown in Fig. \ref{stable2}.

\begin{figure}[htbp]
\centering%
{\includegraphics[scale=0.4]{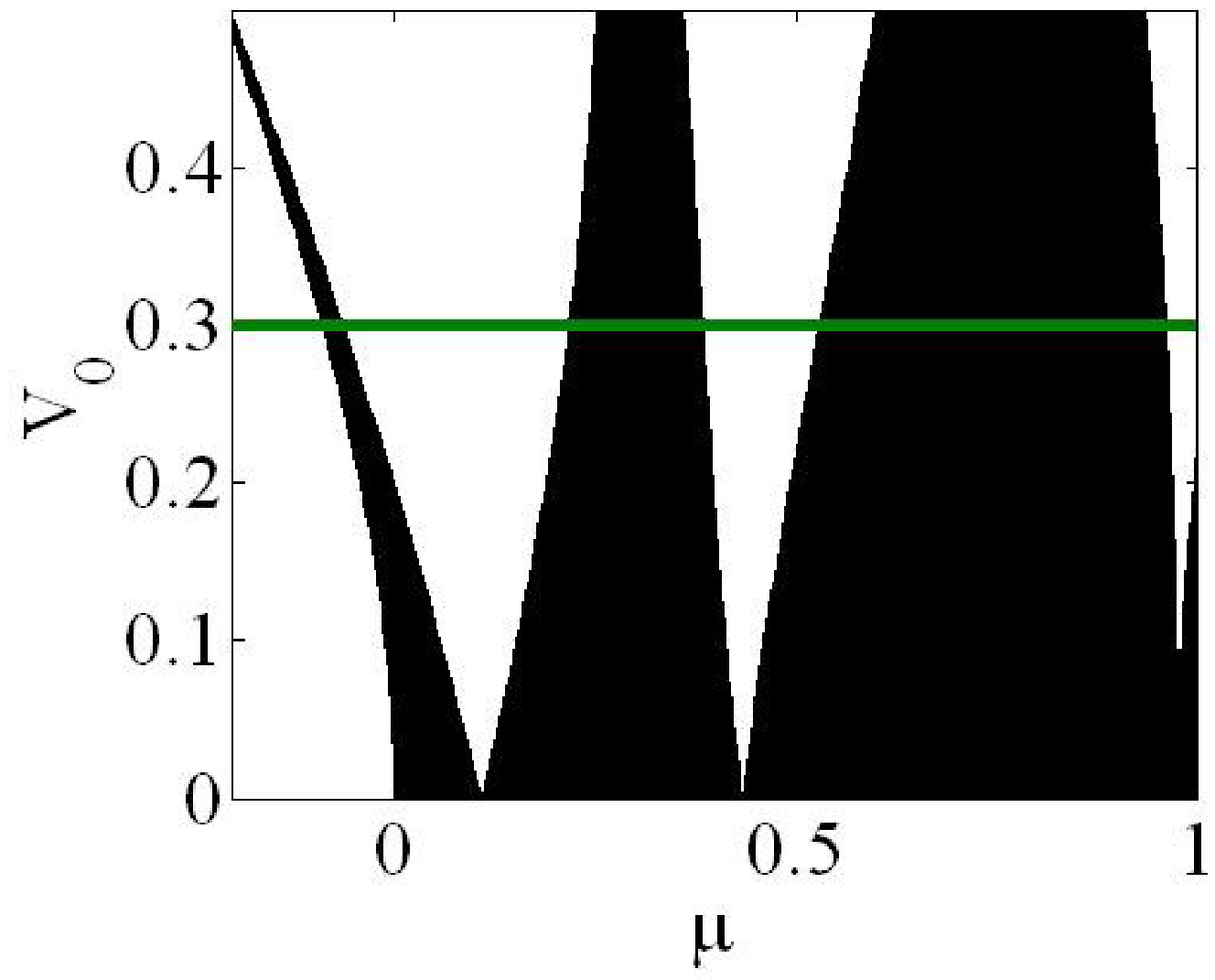}}
\caption{Band structure of the device.}
\label{bandstruc}
\end{figure}

All of above solitons appear in the region of $-0.06<\mu<0.01$, which is very close to the zero of the propagation constant. According to Fig. \ref{bandstruc}, this region belongs to the first bank gap of the periodical structure. For the solution of multi-peak fundamental solitons \cite{Mayteevarunyoo,CLi1,Jwang,CLi2} and the multi-peak dipole solitons with an anti-symmetric profile \cite{YVK4}, it has been report that they can be found in the nonlinear photonic crystals with self-defocusing nonlinearity. And the multi-peak dipole solitons with an asymmetry profile only appear in the surface of the lattice \cite{YVK5,GYin}. However, the asymmetry dipole solitons (ie. Fig \ref{Fig11a}-\ref{Fig13a} and \ref{Fig15a} inside the lattice are observed for the first time though our model. Moreover, the peaks number of the solitons features interesting digital property, which may have potential in optical communication.

\section{CONCLUSION}

Photonic waves propagating under the combination of linear and nonlinear lattice potentials exhibit a number of dynamics \cite{YVK6,Hao}. In this paper, a new type of optical system is built, which juxtaposes three channels (Linear, self-focusing and -defocusing nonlinear) into a period and features the inverted modulation of the linear and nonlinear potentials. The structural inspiration of the medium of this type comes from the recent experimental work in color filter matrix, which use the SU-8 polymer material periodically doped with two types active dyes, which can feature saturable nonlinearity. This model can be described by a nonlinear schr\"{o}dinger equation with two types of saturable nonlinear terms. The numerical simulation finds that some multi-peak fundamental and dipole solitons can stable exist in the first band gap of the lattice. These solitons show some interesting digital properties, which may have potential in optical communication. Furthermore, using properly designed holographic patterns onto the SU-8 polymer, various complex spatial structures of the distribution of the dopant concentration can be photoinduced in the 2D geometry, such as quasicrystals \cite{Freedman,Freedman2}, honeycomb lattices \cite{Peleg,Treidel}, defect lattices \cite{Fedele,YangXY} and ring lattices \cite{WXS,Fleischer2}, etc. Such 2D structures may have their own spectra of potential applications.

G. Chen appreciates the useful discussions with Dr. Yongyao Li. This work is supported by the Science and Technological Program for Dongguan's Higher Education, Science and Research, and Health Care Institutions (No.2011108102025) and the Research Programs of Dongguan University of Technology (No.2010ZQ06).

%\newpage %Just because of unusual number of tables stacked at end
%

\bibliography{apssamp}% Produces the bibliography via BibTeX.

\begin{thebibliography}{99}
\bibitem{Lederer}F. Lederer, G. I. Stegeman, D. N. Christodoulides, G. Assanto, M. Segev, and Y. Silberberg, Discrete solitons in optics, Phys. Rep. {\bf 463} (2008) 1 .
\bibitem{Christodoulides}D. N. Christodoulides, F. Lederer, and Y. Silberberg, Discretizing light behaviour in linear and nonlinear waveguide lattices, Nature (London) {\bf 424} (2003) 817.
\bibitem{ZChen}Z. Chen, M. Segev, D.N. Christodoulides , Optical spatial solitons: historical overview and recent advances, Rep. Prog. Phys. {\bf 75} (2012) 086401.
\bibitem{YVK1}Y. V. Kartashov, V. A. Vysloukh, and L. Torner, Soliton shape and mobility control in optical lattices, Prog. Opt. {\bf 52} (2009) 63.
\bibitem{schwartz}T. Schwartz, G. Bartal, S. Fishman, and M. Segev, Transport and Anderson localization in disordered two-dimensional photonic lattices, Nature(London) {\bf 446} (2007) 52.
\bibitem{Joushaghani}A. Joushaghani, R. Iyer, J. K. S. Poon, J. S. Aitchison, C. M. de Sterke, J. Wan, and
M. M. Dignam, Quasi-Bloch oscillations in curved coupled optical waveguides, Phys. Rev. Lett. {\bf 103} (2009) 143903.
\bibitem{Trompeter}H. Trompeter,W. Krolikowski, D. N. Neshev, A. S. Desyatnikov, A. A. Sukhorukov, Y. S. Kivshar, T. Pertsch, U. Peschel, and F. Lederer, Bloch oscillations and Zener tunneling in two-dimensional photonic lattices, Phys. Rev. Lett. {\bf 96} (2006) 053903.
\bibitem{Eisenberg}H. S. Eisenberg, Y. Silberberg, R. Morandotti, A. R. Boyd, and J. S. Aitchison, Discrete spatial optical solitons in waveguide arrays, Phys. Rev. Lett. {\bf 81} (1998) 3383.
\bibitem{Iwanow}R. Iwanow, R. Schiek, G. I. Stegeman, T. Pertsch, F. Lederer, Y. Min, and W. Sohler, Observation of discrete quadratic solitons, Phys. Rev. Lett. {\bf 93} (2004) 113902.
\bibitem{Fleischer}J. W. Fleischer, M. Segev, N. K. Efremidis, and D. N. Christodoulides, Observation of two-dimensional discrete solitons in optically induced nonlinear photonic lattices, Nature(London) {\bf 422} (2003) 147.
\bibitem{Efremidis}N. K. Efremidis, S. Sears, D. N. Christodoulides, J. W. Fleischer, and M. Segev, Discrete solitons in photorefractive optically induced photonic lattices, Phys. Rev. E {\bf 66} (2002) 046602.
\bibitem{Yongyao1}Y. Li, W. Pang, Y. Chen, Z. Yu, J. Zhou and H. Zhang, Defect-mediated discrete solitons in optically induced photorefractive lattices, Phys. Rev. A {\bf 80} (2009) 043824.
\bibitem{Fratalocchi}A. Fratalocchi, G. Assanto, K. A. Brzdakiewicz, and M. A. Karpierz, Discrete propagation and spatial solitons in nematic liquid crystals, Opt. Lett. {\bf 29} (2004) 1530.
\bibitem{Yingwu}Y. Wu and L. Deng, Ultraslow optical solitons in a cold four-state medium, Phys. Rev. Lett. {\bf 93} (2004) 143904.
\bibitem{GHuang}G. Huang, L. Deng, and M. G. Payne, Dynamics of ultraslow optical solitons in a cold three-state atomic system, Phys. Rev. E {\bf 72} (2005) 016617.
\bibitem{Ghuang2}G. Huang, K. Jiang, M. G. Payne, and L. Deng, Formation and propagation of coupled ultraslow optical soliton pairs in a cold three-state double-$Lambda$ system, Phys. Rev. E {\bf 73} (2006) 056606.
\bibitem{Thong}T. Hong, Spatial weak-light solitons in an electromagnetically induced nonlinear waveguide, Phys. Rev. Lett. {\bf 90} (2003) 183901.
\bibitem{CHuang}C. Hang, G. Huang, and L. Deng, Stable high-dimensional spatial weak-light solitons in a resonant three-state atomic system, Phys. Rev. E {\bf 74} (2006) 046601.
\bibitem{Yongyao4}Y. Li, B. A. Malomed, M. Feng, and J. Zhou Arrayed and checkerboard optical waveguides controlled by the electromagnetically induced transparency, Phys. Rev. A {\bf 82} (2010) 063813.
\bibitem{Weipang}W. Pang, J. Wu, Z. Yuan, Y. Liu and G. Chen, Lattice solitons in optical lattice controlled by electromagnetically induced transparency, J. Phys. Soc. Jpn. {\bf 80} (2011) 113401.
\bibitem{JWu}J. Wu, M. Feng, W. Pang, S. Fu, and Y. Li, The Transmission of quasi-discrete solitons in resonant waveguide arrays activated by the electromagnetically induced transparency, J. Nonlinear Opt. Phys. {\bf 20} (2011) 193.
\bibitem{JGao}J. Gao, H. Li, L. Li, Z. Mai, and G. Chen, Electromagnetically induced quantum lattice soliton, J. Nonlinear Opt. Phys. {\bf 21} (2012) 1250011.
\bibitem{Blaauboer}M. Blaauboer, B. A. Malomed, and G. Kurizki, Spatiotemporally localized multidimensional solitons in self-induced transparency media, Phys. Rev. Lett. {\bf 84} (2000) 1906.
\bibitem{Prineas}J. P. Prineas, J. Y. Zhou, J. Kuhl, H. M. Gibbs, G. Khitrova, S. W. Koch, and A. Knorr, Ultrafast ac Stark effect switching of the active photonic band gap from Bragg-periodic semiconductor quantum wells, Appl. Phys. Lett. {\bf 81} (2002) 4332.
\bibitem{Toader}O. Toader, S. John, and K. Busch, Optical trapping, field enhancement and laser cooling in photonic crystals, Opt. Exp. {\bf 8} (2001) 217.
\bibitem{JYZhou} J. Y. Zhou, H. G. Shao, J. Zhao, X. Yu, and K. S. Wong, Storage and release of femtosecond laser pulses in a resonant photonic crystal, Opt. Lett. {\bf 30} (2005) 1560.
\bibitem{Melnikov}I. V. Mel'nikov and J. S. Aitchison, Aitchison JS. Gap soliton memory in a resonant photonic crystal, Appl. Phys. Lett. {\bf 87} (2005) 201111.
\bibitem{Khomeriki}R. Khomeriki and J. Leon, Driving light pulses with light in two-level media, Phys. Rev. Lett. {\bf 99} (2007) 183601.
\bibitem{Hu1}S. Hu, D. Lu, X. Ma, Q. Guo and W. Hu, Defect solitons supported by nonlocal PT symmetric superlattices, EPL- Europhys. Lett. {\bf 98} (2012) 14006.
\bibitem{Hu2}S. Hu, X. Ma, D. Lu, Y. Zheng, and W. Hu, Defect solitons in parity-time-symmetric optical lattices with nonlocal nonlinearity, Phys. Rev. A {\bf 85} (2012) 043826.
\bibitem{Makris1}K. G. Makris, R. El-Ganainy, D. N. Christodoulides, and Z. H. Musslimani, PT-symmetric optical lattices, Phys. Rev. A {\bf 81} (2010) 063807.
\bibitem{Makris2}K. G. Makris, R. El-Ganainy, and D. N. Christodoulides, Beam Dynamics in PT Symmetric Optical Lattices, Phys. Rev. Lett. {\bf 100} (2008) 103904.
\bibitem{Nixon}S. Nixon, L. Ge, and J. Yang, Stability analysis for solitons in PT-symmetric optical lattices, Phys. Rev. A {\bf 85} (2012) 023822.
\bibitem{Guo}A. Guo, G. J. Salamo, D. Duchesne, R. Morandotti, M. Volatier-Ravat, V. Aimez, G. A. Siviloglou and D. N. Christodoulides, Observation of PT-Symmetry Breaking in Complex Optical Potentials, Phys. Rev. Lett. {\bf 103} (2009) 093902.
\bibitem{Juntao1}J. T. Li and J. Y. Zhou, Nonlinear optical frequency conversion with stopped short light pulses, Opt. Exp. {\bf 14} (2006) 2811.
\bibitem{Yongyao2}Y. Li, W. Pang, S. Fu, and B. A. Malomed, Two-component solitons with a spatially modulated linear coupling: Inverted photonic crystals and fused couplers, Phys. Rev. A {\bf 85} (2012) 053821.
\bibitem{Juntao}J. Li, B. Liang, Y. Liu, P. Zhang, J. Zhou, S. O. Klimonsky, A. S. Slesarev, Y. D. Tretyakov, L. O¡¯Faolain, and T. F. Krauss, Photonic Crystal Formed by the Imaginary Part of the Refractive Index, Adv. Mater. {\bf 22} (2010) 1.
\bibitem{Mfeng} M. Feng, Y. Liu, Yongyao Li, and J. Zhou, Light propagation in a resonantly absorbing waveguide array, Opt. Exp. {\bf 19} (2011) 7222.
\bibitem{Bliang},B. Liang, Y. Liu, L. Song, J. Li, Yongyao Li, J. Zhou, and K. S. Wong, Fabrication of large-size photonic crystals by holographic lithography using a lens array, J. Micromech. Microeng. {\bf 22} (2012) 035013.
\bibitem{Yikunliu}Y. Liu, S. Wang, Y. Li, L. Song, X. Xie, M. Feng, Z. Xiao, S. Deng, J. Zhou, J. Li, K. S. Wong, T. F. Krauss, "Micro-color filter matrix with power transmission exceeding the geometric limit"  Light Science \& Applications, in press.
\bibitem{yongyao3}Y. Li, B. A. Malomed, J. Wu, W. Pang, S. Wang, and J Zhou, Quasicompactons in inverted nonlinear photonic crystals, Phys. Rev. A {\bf 84} (2011) 043839.
\bibitem{YVK2}Y. V. Kartashov, V. A. Vysloukh, and L. Torner, Soliton modes, stability, and drift in optical lattices with spatially modulated nonlinearity, Opt. Lett. {\bf 33} (2008) 1747.
\bibitem{YVK3}Y. V. Kartashov, V. A. Vysloukh, and L. Torner, Power-dependent shaping of vortex solitons in optical lattices with spatially modulated nonlinear refractive index, Opt. Lett. {\bf 33} (2008) 2173.
\bibitem{Yongyao5}Y. Li, B. A. Malomed, M. Feng, and J. Zhou, Double symmetry breaking of solitons in one-dimensional virtual photonic crystals, Phys. Rev. A {\bf 83} (2011) 053832.
\bibitem{JYang}J. Yang, Newton-conjugate-gradient methods for solitary wave computations, J Comput. Phys. {\bf 228} (2009) 7007.
\bibitem{Mayteevarunyoo}T. Mayteevarunyoo, and B. A. Malomed, Solitons in one-dimensional photonic crystals, J. Opt. Soc. Am. B {\bf 25} (2008) 1854.
\bibitem{Jwang}J. Wang, J. Yang, T. J. Alexander, and Yuri S. Kivshar, Truncated-Bloch-wave solitons in optical lattices, Phys. Rev. A {\bf 79} (2009) 043610.
\bibitem{CLi1}C. Li, C. Huang, H. Liu, and L. Dong, Improved sinusoidal phase plate to extend depth of field in incoherent hybrid imaging systems, Opt. Lett. {\bf 37} (2012) 4534.
\bibitem{CLi2}C. Li, H. Liu, and L. Dong, Multi-stable solitons in PT-symmetric optical lattices, Opt. Exp. {\bf 20} (2012) 16823.
\bibitem{YVK4}Y. V. Kartashov, V. A. Vysloukh, and L. Torner, Propagation of solitons in thermal media with periodic nonlinearity, Opt. Lett. {\bf 33} (2008) 1774.
\bibitem{YVK5}Y. V. Kartashov, V. A. Vysloukh, and L. Torner, ultipole surface solitons in thermal media, Opt. Lett. {\bf 34} (2008) 283.
\bibitem{GYin}G. Yin, J. Zheng, X. Yang and L. Dong, Asymmetrical surface soliton trains, Chin. Phys. B {\bf 19} (2010) 044206.
\bibitem{YVK6}Y. V. Kartashov, B. A.Malomed, and L. Torner, Solitons in nonlinear lattices, Rev. Mod. Phys. {\bf 83} (2011) 247.
\bibitem{Hao}R. Hao, Spatial solitons in an optical lattice with varying diffraction and spatially inhomogeneous nonlinearity, Opt. Laser Technol. {\bf 43} (2011) 25.
\bibitem{Freedman}B. Freedman, G. Bartal, M. Segev. R. Lifshitz, D. N. Christodoulides, and J.W. Fleischer, Wave and defect dynamics in nonlinear photonic quasicrystals, Nature (London) {\bf 440} (2006) 1166.
\bibitem{Freedman2}B. Freedman, R. Lishitz, J. W. Fleischer, and M. Segev, Phason dynamics in nonlinear photonic quasicrystals, Nature Materials (London) {\bf 6} (2007) 776.
\bibitem{Peleg} O. Peleg, G. Bartal, B. Freedman, O. Manela, M. Segev, and D. N. Christodoulides, Conical diffraction and gap solitons in honeycomb photonic lattices, Phys. Rev. Lett. {\bf 98} (2007) 103901.
\bibitem{Fedele}F. Fedele, J. Yang, and Z. Chen, Defect modes in one-dimensional photonic lattices, Opt. Lett. {\bf 30} (2005) 1506.
\bibitem{YangXY}X. Yang, J. Zheng, and L. Dong, Spatial solitons in photonic lattices with large-scale defects, Chin. Phys. B, {\bf 20} (2011) 034208.
\bibitem{Treidel}O. B. Treidel, O. Peleg, and M. Segev, Symmetry breaking in honeycomb photonic lattices, Opt. Lett. {\bf 33} (2009) 2251.
\bibitem{WXS} X. Wang, Z. Chen, and P. G. Kevrekidis, Observation of discrete solitons and soliton rotation in optically induced periodic ring lattices, Phys. Rev. Lett. {\bf 96} (2006) 083904.
\bibitem{Fleischer2}J. W. Fleischer, G. Bartal, O. Cohen, O. Manela, M. Segev, J. Hudock, and D. N. Christodoulides, Observation of vortex-ring discrete solitons in 2D photonic lattices, Phys. Rev. Lett. {\bf 92} (2004) 123904.
\end{thebibliography}

\end{document}